# Experimental demonstration of near-infrared epsilon-near-zero multilayer metamaterial slabs

Xiaodong Yang,[1,*] Changyu Hu,[1] Huixu Deng,[1] Daniel Rosenmann,[2] David A. Czaplewski,[2] and Jie Gao[1,3]

[1]*Department of Mechanical and Aerospace Engineering, Missouri University of Science and Technology, Rolla, MO 65409, USA*
[2]*Center for Nanoscale Materials, Argonne National Laboratory, Argonne, IL 60439, USA*
[3]*gaojie@mst.edu*, [*]*yangxia@mst.edu*

**Abstract:** Near-infrared epsilon-near-zero (ENZ) metamaterial slabs based on silver-germanium (Ag-Ge) multilayers are experimentally demonstrated. Transmission, reflection and absorption spectra are characterized and used to determine the complex refractive indices and the effective permittivities of the ENZ metamaterial slabs, which match the results obtained from both the numerical simulations and the optical nonlocalities analysis. A rapid post-annealing process is used to reduce the collision frequency of silver and therefore decrease the optical absorption loss of multilayer metamaterial slabs. Furthermore, multilayer grating structures are studied to enhance the optical transmission and also tune the location of ENZ wavelength. The demonstrated near-infrared ENZ multilayer metamaterial slabs are important for realizing many exotic applications, such as phase front shaping and engineering of photonic density of states.

## 1. Introduction

Metamaterials are artificially engineered structures designed to achieve extraordinary electromagnetic properties that are not available with natural materials [1, 2]. Recently, epsilon-near-zero (ENZ) metamaterials have been studied to realize anomalous electromagnetic properties in microwave and optical frequency ranges. ENZ metamaterials can be used to squeeze and tunnel electromagnetic energy through narrow channels due to the extremely large wavelength they support [3-6]. The near-zero phase variation of electromagnetic wave propagating inside ENZ metamaterials suggests applications of directive emission and phase front shaping [7, 8]. Self-collimation and focusing effects are demonstrated in zero-average index metamaterials supporting defect modes [9]. The inherent negative polarizability of the ENZ metamaterials also enables potential applications in electromagnetic transparency and invisible cloaking [10-13]. In addition, displacement current insulation in optical nanocircuits and subwavelength image transporting have also been designed with ENZ metamaterials [14, 15]. ENZ metamaterials can also significantly enhance optical nonlinearities [16] and photonic density of states [17].

Natural ENZ materials are readily available as noble metals, doped semiconductors [18], polar dielectrics [15], and transparent conducting oxides (TCOs) [19]. The permittivity can be characterized by a Drude or a Drude-Lorentz model, and the near-zero permittivity is obtained when the frequency approaches the plasma frequency. Since noble metals exhibit plasma frequencies in the ultraviolet region, inclusions of noble metals can be embedded in a dielectric host medium to synthesize ENZ materials at visible or near-infrared frequencies, based on the effective medium theory [20]. ENZ properties can be realized by utilizing metal coated waveguides at the cutoff frequency [17, 21], arrays of silver or gold nanowires grown in porous alumina templates [22, 23], and metal-dielectric multilayer structures [24-26]. Multilayer metamaterials have been explored to realize extraordinary optical functionalities such as negative refraction [27, 28], hyperlenses for sub-wavelength imaging [29], and indefinite metamaterial cavities [30]. In this paper, ENZ metamaterial slabs at near-infrared frequencies based on silver-germanium (Ag-Ge) multilayers are fabricated and characterized to study their unique optical properties and determine the ENZ wavelengths. The complex refractive indices and effective permittivities of the ENZ metamaterial slabs are retrieved from the measured transmission and reflection spectra, which agree with the results obtained from finite element method (FEM) numerical simulations and optical nonlocalities analysis. A rapid post-annealing process is used to reduce the collision frequency of silver and therefore decrease the absorption loss of ENZ metamaterial slabs. Moreover, grating structures based on multilayer metamaterial slabs are studied in order to enhance the optical transmission and also tune the ENZ wavelength. The demonstrated near-infrared ENZ multilayer metamaterial slabs will provide new opportunities in studying intriguing optical phenomena, such as phase front shaping [8], optical nonlinearities enhancement [16], spontaneous emission control [31], and active metamaterials [32].

## 2. Optical nonlocalities analysis of metal-dielectric multilayers

Figure 1(a) shows the schematic of Ag-Ge multilayer structures with layer permittivities of $\varepsilon_1$, $\varepsilon_2$ and layer thicknesses of $d_1$, $d_2$, where 1 and 2 represent Ag and Ge respectively. The permittivity of Ag is described by the Drude model $\varepsilon_1 = \varepsilon_\infty - \omega_p^2/(\omega^2 - i\omega\gamma_p)$, where the background dielectric constant is $\varepsilon_\infty = 5.0$, the plasma frequency is $\omega_p = 1.38 \times 10^{16}$ rad/s, and the collision frequency $\gamma_p$ is equal to $\gamma_0 = 5.07 \times 10^{13}$ rad/s for bulk Ag [33]. For Ag thin films, $\gamma_p$ increases due to the electron scattering at the metal boundaries. The permittivity of Ge is obtained from the refractive index database [34]. It has been shown that metal-dielectric multilayer structures have strong optical nonlocalities [35, 36], where the permittivity components are functions of both frequencies and wave vectors. Multilayer structures can be considered as a one-dimensional photonic crystal along the $x$-direction. According to the transfer matrix method, the dispersion relation for eigenmodes with transverse-magnetic (TM) polarization is derived as,

$$\cos(k_x(d_1+d_2)) = \cos(k_x^{(1)}d_1)\cos(k_x^{(2)}d_2) - \frac{1}{2}\left(\frac{\varepsilon_1 k_x^{(2)}}{\varepsilon_2 k_x^{(1)}} + \frac{\varepsilon_2 k_x^{(1)}}{\varepsilon_1 k_x^{(2)}}\right)\sin(k_x^{(1)}d_1)\sin(k_x^{(2)}d_2) \qquad (1)$$

where $k_x^{(1,2)} = \sqrt{k_0^2 \varepsilon_{1,2} - k_y^2}$ are $x$ components of wave vectors in metal and dielectric layers, and $k_0 = \omega/c$ is the wave vector in free space. Equation (1) can be written in the form of

$$\frac{k_x^2}{\varepsilon_y^{\text{eff}}} + \frac{k_y^2}{\varepsilon_x^{\text{eff}}} = k_0^2 \qquad (2)$$

with the effective permittivity components of $\varepsilon_x^{\text{eff}}(\omega,k)$ and $\varepsilon_y^{\text{eff}}(\omega,k)$ including the nonlocal effects. For TM polarized light at normal incidence with $k_y = 0$, $\varepsilon_y^{\text{eff}} = k_x^2/k_0^2$ represents the effective dielectric constant of the metamaterial slab and it can be solved from Eq. (1) as,

$$\varepsilon_y^{\text{eff}} = \frac{\arccos^2\left[\cos\left(\sqrt{\varepsilon_1}k_0 d_1\right)\cos\left(\sqrt{\varepsilon_2}k_0 d_2\right) - \frac{1}{2}\left(\sqrt{\varepsilon_1/\varepsilon_2} + \sqrt{\varepsilon_2/\varepsilon_1}\right)\sin\left(\sqrt{\varepsilon_1}k_0 d_1\right)\sin(\sqrt{\varepsilon_2}k_0 d_2)\right]}{k_0^2(d_1+d_2)^2} \qquad (3)$$

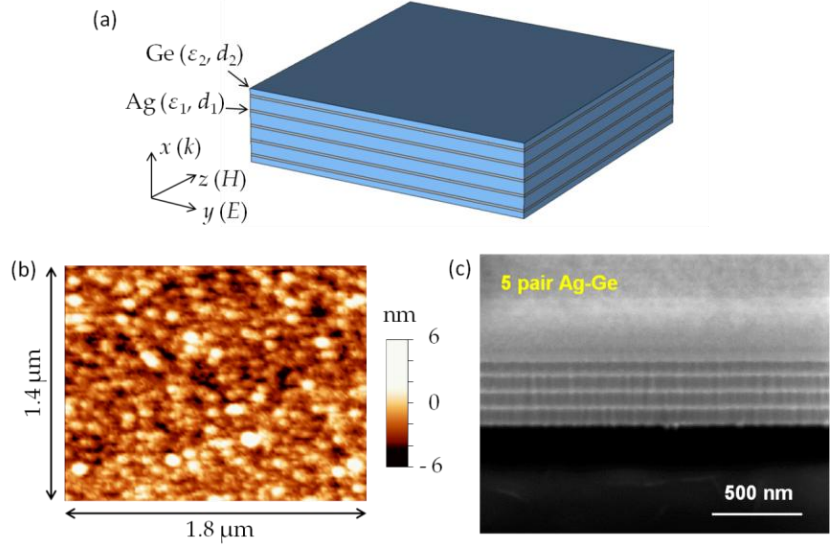

**Fig. 1. (a)** Schematic of Ag-Ge multilayer metamaterial slabs, with layer permittivities of $\varepsilon_1$, $\varepsilon_2$ and layer thicknesses of $d_1$, $d_2$, where 1 and 2 represent Ag and Ge respectively. The incident light is TM polarized propagating along the $x$ direction with the components of $E_y$ and $H_z$. **(b)** A representative AFM picture showing the surface roughness distribution for the fabricated Ag-Ge multilayer metamaterial slabs. The RMS roughness is 1.6 nm. **(c)** A SEM picture of the cross section of fabricated Ag-Ge multilayer metamaterial slabs made of 5 pairs of 15 nm Ag and 85 nm Ge multilayers. The bright and dark stripes correspond to Ag and Ge layers, respectively.

## 3. Fabrication and characterization of ENZ metamaterial slabs

The Ag-Ge multilayer metamaterial slabs consist of 5 pairs of 15 nm Ag and 85 nm Ge evaporated on quartz substrates. The films were produced in a Lesker PVD-250 e-beam evaporator system, at a base pressure of $8 \times 10^{-9}$ Torr. The quartz substrates were rotated at 20 rpm and kept at room temperature by means of a chilled water-cooling stage. Ag and Ge films were deposited, through a QCM feedback control, at a rate of 0.02 nm/s. Both Ag and Ge are individually deposited on quartz substrates first to calibrate and optimize the deposition parameters for the electron-beam evaporation system. Then 5 pairs of Ag-Ge multilayers are fabricated as shown in Fig. 1(a), where a half layer of Ge with thickness 42.5 nm is deposited as top and bottom layers in order to achieve symmetrical multilayer structures and also to protect the Ag from oxidizing in air. According to the atomic force microscope (AFM) measurements, the root mean squared (RMS) roughness of the top surface of the fabricated multilayers is only 1.6 nm. Figure 1(b) shows a representative AFM picture of the surface roughness distribution for the fabricated multilayers. Figure 1(c) shows a scanning electron microscope (SEM) picture of the cross section of the fabricated multilayers, created by ion milling using a focused ion beam (FIB) system (Helios Nanolab 600). Each deposited thin layer can be clearly seen, where the bright and dark stripes correspond to Ag layers and Ge layers, respectively.

In order to characterize the optical properties of the fabricated ENZ metamaterial slabs, optical transmission ($T$) and reflection ($R$) spectra are measured with a Fourier transform infrared spectroscopy (FTIR, Nicolet 6700) and an infrared microscope. The transmission spectrum is normalized with a quartz wafer, and the reflection spectrum is normalized with a silver mirror. Optical absorption ($A$) spectrum is then obtained from $A = 1 - T - R$. Figure 2(a) plots the measured $T$, $R$ and $A$ for the metamaterial slab, together with the FEM simulation results. Figure 2(b) gives the complex refractive indices ($n + ik$) determined from the measured transmission, $T$, and reflection, $R$, using an approximated approach based on incoherent interference (Method 1) [37, 38],

$$k = \frac{-\lambda}{4\pi t} \ln\left\{\frac{[T^2-(1-R)^2]+\{[T^2-(1-R)^2]^2+4T^2\}^{1/2}]}{2T}\right\} \tag{4}$$

$$n_\pm = \frac{(1+R_{as})}{(1-R_{as})} \pm \left[\frac{4R_{as}}{(1-R_{as})^2} - k^2\right]^{1/2} \tag{5}$$

where $n_+$ is selected when $n \geq 1$ and $n_-$ is selected when $n < 1$. The surface reflectance between air and the metamaterial slab, $R_{as}$, is

$$R_{as} = \frac{R}{1+\left[[T^2-(1-R)^2]+\{[T^2-(1-R)^2]^2+4T^2\}^{1/2}\right]/2} \tag{6}$$

The complex refractive indices ($n + ik$) are also obtained from the FEM simulated $S$ parameters of reflectance, $S_{11}$, and transmittance, $S_{21}$, (including both amplitude and phase), according to the algorithm to retrieve the constitutive effective parameters of metamaterials [39] (Method 2). Note that another graphical retrieval method with phase unwrapping techniques can also be used to accurately predict the effective parameters of bulk metamaterials from a single layer of unit cells [40]. In addition, the values of $n$ and $k$ calculated from the nonlocal dispersion relation based on Eq. (3) are plotted for comparison (Method 3). Note that the effective permittivity, $\varepsilon_y^{\text{eff}} = (n+ik)^2$, since the permeability $\mu = 1$ for the multilayer metamaterial slabs. As shown in Fig. 2(b), the complex refractive indices calculated from Method 2 and Method 3 agree with each other. The refractive indices determined from the measured transmission and reflection without the phase information in Method 1 can reproduce the same trend as the calculation results from Method 2 and Method 3. *Nota bene*, the ENZ wavelength determined from all three methods overlaps at 1.635 μm, where $n$ is equal to $k$. Figure 2(c) gives the effective permittivity obtained from the relation $\varepsilon_y^{\text{eff}} = (n+ik)^2$, according to the above three methods. The location of the ENZ wavelength where $\text{Re}(\varepsilon_y^{\text{eff}}) = 0$ is predicted accurately with Method 1, although the slope of $\text{Re}(\varepsilon_y^{\text{eff}})$ is slightly different from the calculated values from Method 2 and Method 3. The $\text{Im}(\varepsilon_y^{\text{eff}})$ shows a good trend when compared to the calculation results. The ENZ wavelength can be determined accurately for Ag-Ge multilayer metamaterial slabs based on the measured transmission and reflection spectra.

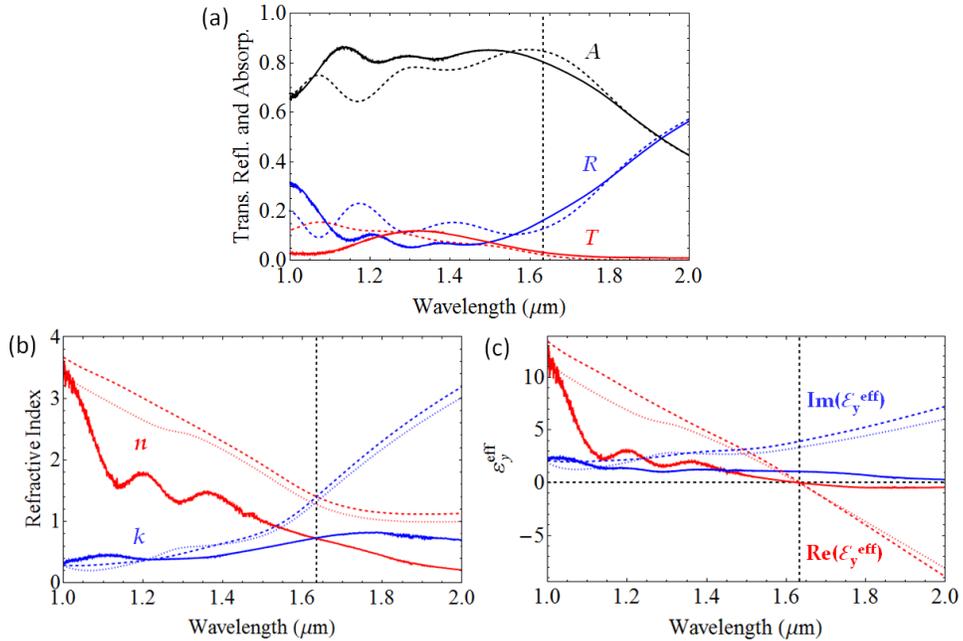

**Fig. 2. (a)** Transmission (*T*), reflection (*R*), and absorption (*A*) spectra for the 5 pairs of Ag-Ge multilayers. The solid curves represent the measured data, while the dashed curves show the FEM simulation results. **(b)** The retrieved complex refractive indices *n* and *k* for the metamaterial slab. The solid curves represent the values determined from the FTIR measured transmission and reflection (Method 1). The dotted curves represent the FEM simulation retrieved values (Method 2), and the dashed curves show the results of optical nonlocality analysis (Method 3). **(c)** Real and imaginary parts of the effective permittivity $\varepsilon_y^{eff}$ obtained from Methods 1, 2, and 3.

The collision frequency of the Ag thin film, $\gamma_p$, is usually several times higher than that of bulk silver, $\gamma_0$, due to the additional surface scattering and grain boundary effects in Ag thin films, when the size of metal structure is smaller than the mean free path of the electrons. Such quantum size effects can be included in the collision frequency as $\gamma_p = \gamma_0 + A_0 \frac{v_F}{r}$ [41, 42], where $v_F$ is the Fermi velocity of Ag, *r* is the metal particle radius and $A_0$ is a dimensionless constant depending on the scattering process. The theoretical collision frequency for a 15 nm thick Ag film can be calculated with the grain size of *r* = 15 nm, $A_0$ = 0.25, $v_F$ = 1.39×10$^6$ m/s [41, 42], which results in $\gamma_p$ = 3.8 $\gamma_0$. However, the collision frequency used in the FEM simulations of Fig. 2(a) is 6.4 $\gamma_0$ in order to fit the measured spectra. Here a rapid post-annealing treatment [42] is used to increase the restricted electron mean free path in the Ag thin film and therefore reduce the collision frequency close to the theoretical value. The multilayer metamaterial slab is treated with the annealing process by varying the temperature from 160 °C to 280 °C with time duration from 3 minutes to 7 minutes. The results in Fig. 3 show that a 3-minute annealing treatment at 180 °C can reduce the absorption loss and increase the transmission and reflection of metamaterial slabs, resulting in a lower collision frequency of 5.4 $\gamma_0$ according to the FEM simulations. The retrieved complex refractive indices (*n* + *ik*) are plotted in Fig. 3(d), indicating a lower *k* value after the annealing treatment.

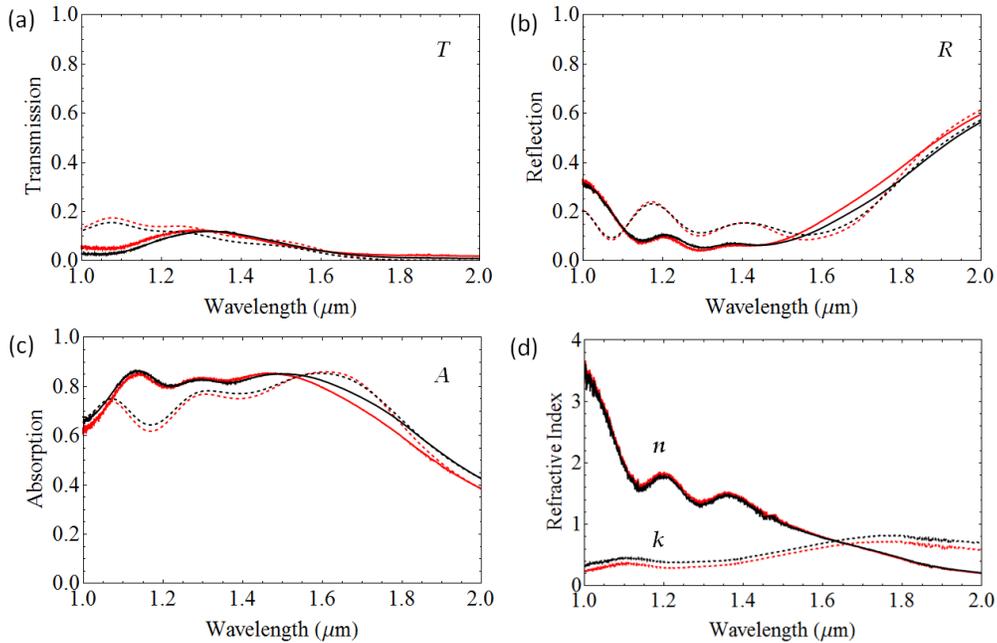

**Fig. 3. (a)** Transmission, **(b)** reflection, and **(c)** absorption spectra for the multilayer metamaterial slabs before and after the annealing treatment at 180 °C for 3 minutes. The solid curves represent the FTIR measured data, while the dashed curves show the FEM simulation results. Black colored curves give data before the annealing and red colored curves show results after the annealing. **(d)** The retrieved complex refractive indices *n* and *k* from the FTIR measured transmission and reflection before and after the annealing. The solid curves represent *n* and the dashed curves show *k*. Black colored curves give data before the annealing and red colored curves show results after the annealing.

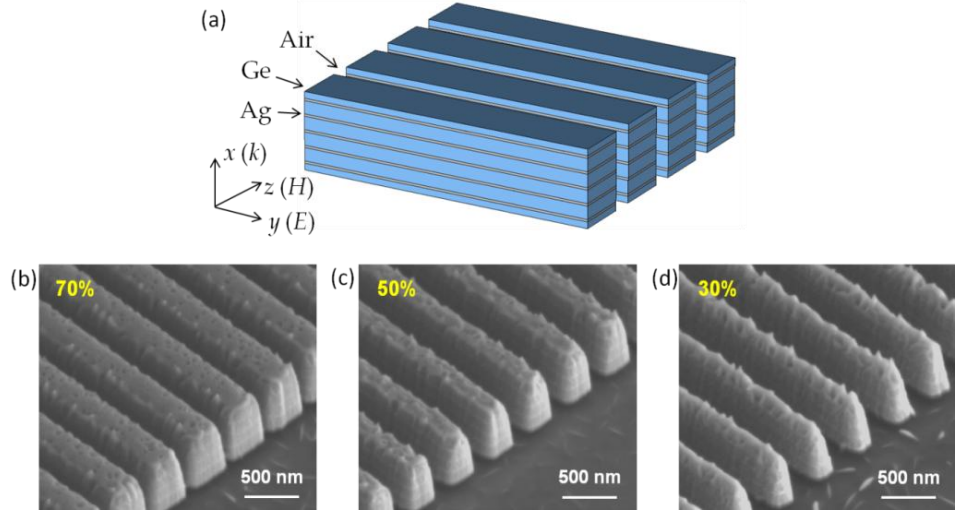

**Fig. 4. (a)** Schematic of multilayer grating structures along the *y* direction, which mix the multilayer and air together. The incident light is TM polarized propagating along the *x* direction with the components of $E_y$ and $H_z$. SEM pictures of the cross sections of fabricated multilayer grating structures with the period of 500 nm and the duty cycles of **(b)** 70%, **(c)** 50% and **(d)** 30%, respectively.

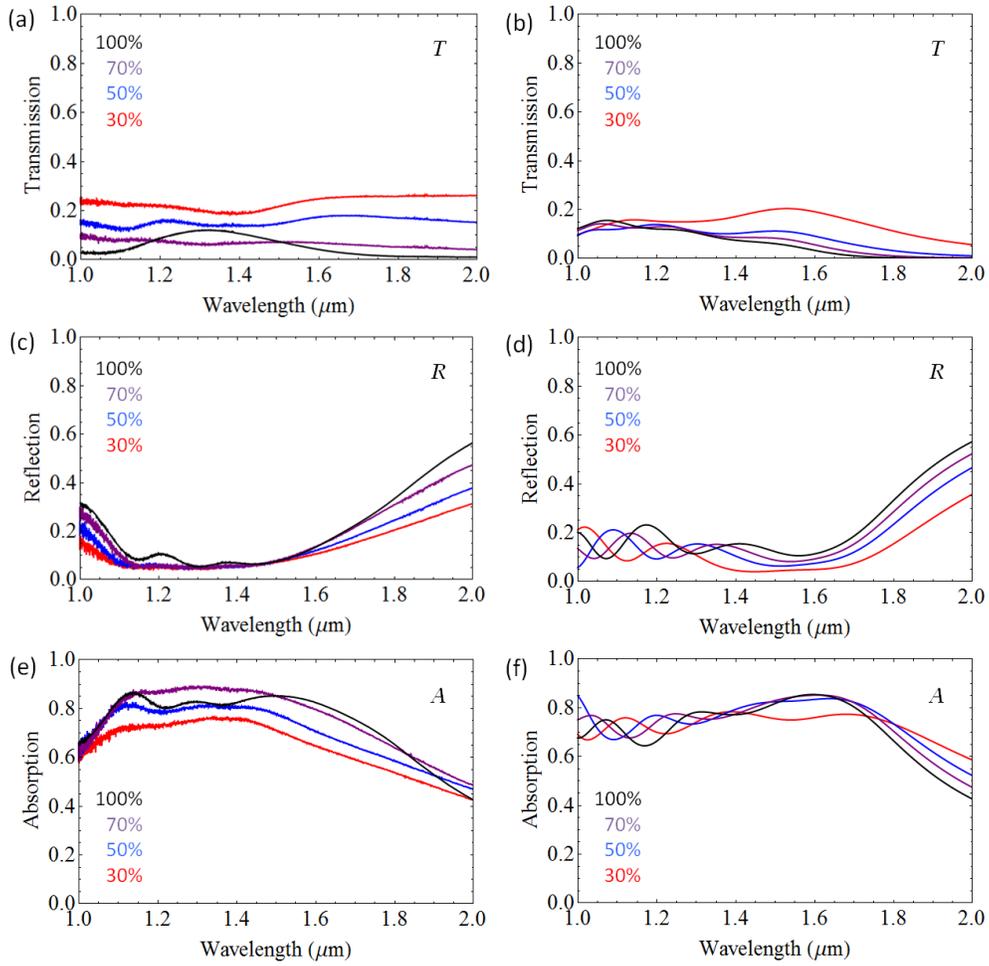

**Fig. 5.** FTIR measured **(a)** transmission, **(c)** reflection and **(e)** absorption spectra for the fabricated multilayer grating structures with different duty cycles. FEM simulated **(b)** transmission, **(d)** reflection and **(f)** absorption spectra are shown to agree with the measured data.

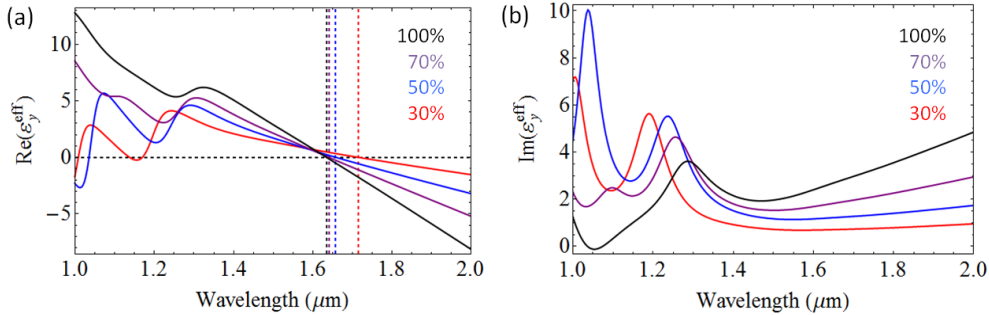

**Fig. 6.** Real and imaginary parts of the effective permittivity $\varepsilon_y^{eff}$ retrieved from the FEM simulations (Method 2) for multilayer grating structures with different duty cycles. The vertical dashed lines in (a) show the locations of ENZ wavelengths.

In order to enhance the optical transmission and also tune the ENZ wavelength, grating structures were fabricated with FIB along the *y* direction on the multilayer metamaterial slabs. Figure 4(a) gives the schematic of the multilayer grating structure, where the grating period is fixed and the duty cycle of the grating structure varies. Figures 4(b), 4(c) and 4(d) show the SEM pictures of the multilayer grating structures with a period of 500 nm and duty cycles of 70%, 50% and 30%. Figures 5(a), 5(c) and 5(e) plot the FTIR measured optical transmission, reflection, and absorption spectra for the fabricated multilayer grating structures. As the duty cycle gets smaller, higher transmission and lower reflection and absorption are achieved due to the fact that the larger filling ratio of air in the grating structure will reduce the amount of Ag and therefore result in a lower optical loss. Figures 5(b), 5(d) and 5(f) give the FEM simulated results for the transmission, reflection and absorption spectra for multilayer grating structures with different duty cycles, which agree with the measured data. The complex refractive indices ($n + ik$) are then retrieved from the simulated *S* parameters of reflectance $S_{11}$ and transmittance $S_{21}$ (Method 2), and the effective permittivity is plotted in Fig. 6 according to $\varepsilon_y^{\text{eff}} = (n + ik)^2$. It is shown that the ENZ wavelength where $\text{Re}(\varepsilon_y^{\text{eff}}) = 0$ is red-shifted and the dispersion near the ENZ wavelength becomes flatter as the duty cycle reduces, which is due to the increase of the filling ratio of air in the grating structure. Since the magnitude of the real part of the permittivity of Ag is much larger than that of Ge, the increase of the air percentage will result in more reduction on the magnitude of the effective permittivity of the Ag-air layer than that of the Ge-air layer, leading to a red-shifted ENZ wavelength to compensate this reduction. The flatter dispersion near the ENZ wavelength is due to the increase of the percentage of non-dispersive air. The corresponding ENZ wavelength for the duty cycle of 100%, 70%, 50% and 30% is 1.635 µm, 1.640 µm, 1.658 µm, 1.715 µm, respectively. At the same time, the $\text{Im}(\varepsilon_y^{\text{eff}})$ gets lower so that the absorption loss of metamaterial slab is reduced significantly due to the larger percentage of lossless air.

## 4. Conclusion

Ag-Ge multilayer ENZ metamaterial slabs at near-infrared frequencies have been experimentally demonstrated. The complex refractive indices and the effective permittivities of the multilayer metamaterial slabs are determined from the measured transmission and reflection spectra, which match the retrieved values from numerical simulations and the results obtained from optical nonlocalities analysis. Moreover, a rapid post-annealing process is used to reduce the collision frequency of silver and decrease the optical absorption of multilayer metamaterial slabs. Furthermore, multilayer grating structures are designed to enhance the optical transmission and also tune the location of the ENZ wavelength. The demonstrated near-infrared ENZ multilayer metamaterial slabs will enable many exciting applications, such as phase front shaping, optical nonlinearities enhancement, photonic density of states engineering, and active optical metamaterials.

## Acknowledgments

This work was partially supported by the Intelligent Systems Center, the Energy Research and Development Center and the Materials Research Center at Missouri S&T, and the University of Missouri Research Board. The use of the Center for Nanoscale Materials was supported by the U. S. Department of Energy, Office of Science, Office of Basic Energy Sciences, under Contract No. DE-AC02-06CH11357. The authors acknowledge L. Sun for his useful discussions about this work.